\documentclass[pre,twocolumn,usletter,superscriptaddress]{revtex4}

\usepackage{amsmath}
\usepackage{amsfonts}
\usepackage{amssymb}

\usepackage [ansinew] {inputenc}
\usepackage{graphicx}
\usepackage{natbib}

\begin{document}
\title{The $p\lambda n$ fractal decomposition: \\ Nontrivial partitions of conserved physical quantities}

\author{Vladimir Garc\'{\i}a-Morales}

\affiliation{Departament de Termodin\`amica, Universitat de Val\`encia, E-46100 Burjassot, Spain}
\email{garmovla@uv.es}




\begin{abstract}
\noindent A mathematical method for constructing fractal curves and surfaces, termed the $p\lambda n$ fractal decomposition, is presented. It allows any function to be split into a finite set of fractal discontinuous functions whose sum is equal everywhere to the original function. Thus, the method is specially suited for constructing families of fractal objects arising from a conserved physical quantity, the decomposition yielding an exact partition of the quantity in question. Most prominent classes of examples are provided by Hamiltonians and partition functions of statistical ensembles: By using this method, any such function can be decomposed in the ordinary sum of a specified number of terms (generally fractal functions), the decomposition being both exact and valid everywhere on the domain of the function. 
 ~\\

 \end{abstract}
\maketitle

\section{Introduction}

A fractal is a geometrical structure with non-integer dimension and the property of self-similarity \cite{Mandelbrot, Barnsley, Feder, PeitgenBOOK}. Fractals abound in present day applications and are being used to describe a variety of phenomena in nature, ranging from nanoclusters \cite{Moriarty} and semiconductors \cite{Richardella} to galaxies \cite{Martinez}. Fractals are important in the statistical mechanics of complex systems \cite{Tsallis, Tsallis2, Hoover} and quantum mechanics \cite{FeynmanHibbs, Kroeger, Suzuki1, Suzuki2, Laskin}. Mathematically, they can arise from a variety of processes: iterated function systems \cite{PeitgenBOOK, Barnsley}, strange attractors of chaotic dynamical systems \cite{Ott}, critical phenomena \cite{SornetteBOOK}, cellular automata \cite{Wolfram, VGM1, VGM2, VGM3}, substitution systems \cite{VGM4}, bitwise arithmetic \cite{arxivCSo}, and arithmetic sequences \cite{Bondarenko}. Recently, a construction has been presented in terms of the hierarchical structure of a coin division problem \cite{Yamamoto, Yamamoto2}.  

In this article we present a new construction of a wide variety of fractal structures motivated by a general mathematical problem of interest to Physics that is easy to state and for which we provide a solution: \emph{To find whether if given an arbitrary Hamiltonian $H$ is it possible to split it into the ordinary sum of a given finite number $\lambda \in \mathbb{N}$ of (generally discontinuous) functions, bringing the Hamiltonian to the following form}
\begin{equation}
H=\sum_{n=0}^{\lambda-1}H_{n} \label{ispos}
\end{equation}
We call this problem the \emph{partitioning} of the Hamiltonian. With this problem we exclude, of course, the trivial partitioning that results from dividing the Hamiltonian $H$ by $\lambda$  or by forming any linear combination of rational functions of $H$. We ask whether is it \emph{always} possible to find $\lambda$ \emph{nontrivial} partitions of $H$ \emph{regardless} of the form of $H$ (of course, the partitions $H_{n}$ should themselves depend on $H$ in a nontrivial way). Since we strive to give a general, constructive answer, we must renounce to demand smoothness (continuity and/or differentiability) of the parts $H_{n}$.

The answer to this question is, surprisingly, affirmative and we provide an explicit, general method, to construct each part $H_{n}$ given $H$ and $\lambda$. The partition is valid and exact everywhere on the energy surface and, by construction, the $H_{n}$'s generally possess fractal features. Our approach makes use of a digit function that we have recently introduced in a new formulation of classical and quantum mechanics \cite{QUANTUM} as well as elementary facts of abstract algebra (finite group theory) \cite{Gallian}. 
In the following section we first introduce the digit function elucidating some of its mathematical properties that we shall then use. We explicitly prove that the digit function induces a cyclic group structure on a finite set of integers. We then proceed to establish our main result, which is contained in what we term the $p\lambda n$ fractal decomposition: We prove that such a decomposition satisfies Eq. (\ref{ispos}) thus solving the above problem affirmatively.  We then extend these results to complex-valued quantities and prove a theorem that generalizes the method, significantly extending its domain of applicability. We finish with a short discussion of the mathematical results, sketching their connection to statistical mechanics and showing how they constitute a number-theoretic formulation of the latter.

\section{The digit function and its basic properties}

Let $p>1$ be a natural number. Every real number $x$ can be expanded in powers of $p$ as
\begin{equation}
x=a_{N}p^{N}+a_{N-1}p^{N-1}+\ldots+a_{0}+a_{-1}p^{-1}+a_{-2}p^{-2}+\ldots \label{bexpareal}
\end{equation}
where the $a_{k}$'s are the so-called digits of the expansion and are either all positive or all negative, with absolute values being integers in the interval $[0,p-1]$.  Eq. (\ref{bexpareal}) is the representation of the real number $x$ in base $p$. We shall call $p$ the \emph{radix} to avoid confusion with other uses of the word 'base' in physics. Eq. (\ref{bexpareal}) can thus be written as 
\begin{equation}
x=\text{sign}(x)\sum_{k=-\infty}^{\lfloor \log_{p}|x| \rfloor}|a_{k}|p^{k}
\equiv \text{sign}(x)\sum_{k=-\infty}^{\lfloor \log_{p}|x| \rfloor}\mathbf{d}_{p}(k,|x|) p^{k}
 \label{idenreal}
\end{equation}
where the upper bound in the sum gives the total number of integer digits $N$ as $N=\lfloor \log_{p}|x| \rfloor+1$. Above, the digit function $\mathbf{d}_{p}(k,|x|)\equiv |a_{k}|$ of the real number $x$ has been defined through the coefficients $|a_{k}|$. The digit function is explicitly given by \cite{QUANTUM}
\begin{equation}
\mathbf{d}_{p}(k,x) = \left \lfloor \frac{x}{p^{k}} \right \rfloor-p\left \lfloor \frac{x}{p^{k+1}} \right \rfloor   \label{cucuAreal}
\end{equation}
where $\lfloor \ldots \rfloor$ denote the closest lower integer (floor function) \cite{Graham, KnuthII}. To see this note that, from Eq. (\ref{bexpareal})  (we consider $x$ nonnegative for simplicity)
\begin{equation}
\left \lfloor \frac{x}{p^{k}} \right \rfloor=a_{N}p^{N-k}+a_{N-1}p^{N-k-1}+\ldots +a_{k+1}p+a_{k} 
\end{equation}
 From Eq. (\ref{bexpareal}) we have, as well,
\begin{equation}
p\left \lfloor \frac{x}{p^{k+1}} \right \rfloor=a_{N}p^{N-k}+a_{N-1}p^{N-k-1}+\ldots +a_{k+1}p 
\end{equation}
whence, by subtracting both equations, we obtain Eq. (\ref{cucuAreal}). 

\begin{figure*}
\includegraphics[width=0.7\textwidth]{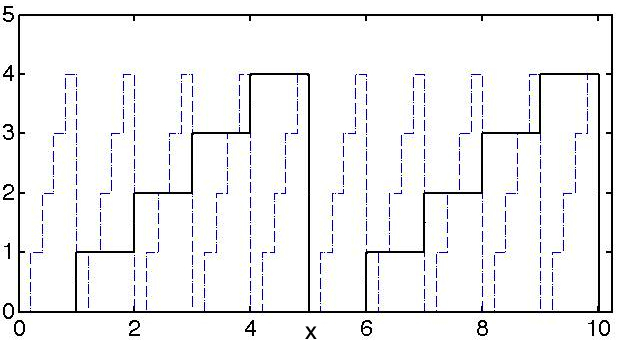}
\caption{The digit function $\mathbf{d}_{5}(0,x)$ (continuous curve) and $\mathbf{d}_{5}(-1,x)$ (dashed curve)  } \label{dig}
\end{figure*}


The digit function $\mathbf{d}_{p}(k,x)$ is plotted in Fig. \ref{dig} for $p=5$ and $k=0$ (continuous curve) and $k=-1$ (dashed curve). It is a staircase of $p$ levels taking discrete integer values between $0$ and $p-1$. Each time that $x$ is divisible by $p^{k+1}$ the ascent of the staircase is broken and the level is  set again to zero and a new staircase begins. We note that, in Eq. (\ref{idenreal}) we have, asymptotically
\begin{eqnarray}
\lim_{p\to \infty}x&=&\lim_{p\to \infty}\left[\text{sign}(x)\sum_{k=-\infty}^{\lfloor \log_{p}|x| \rfloor}\mathbf{d}_{p}(k,|x|) p^{k}\right] \nonumber \\
&\sim&  \text{sign}(x)\mathbf{d}_{p}(\lfloor \log_{p}|x| \rfloor,|x|) p^{\lfloor \log_{p}|x| \rfloor} 
 \label{idenasi}
\end{eqnarray}
or, if the limit is strictly taken 
\begin{equation}
\lim_{p\to \infty}x=\text{sign}(x)\mathbf{d}_{\infty}(0,|x|) \label{thermolim}
\end{equation}
We note that in \cite{QUANTUM} $x$ was taken nonnegative throughout (and hence $|x|=x$ and $\text{sign}(x)=1$ there) and that the digit function as defined $\mathbf{d}_{p}(k,|x|)$ in \cite{QUANTUM} corresponds to $\mathbf{d}_{p}(k-1,|x|)$ here. We make here this notational adjustment to stay more in tune with standard mathematical textbooks \cite{Andrews} where the radix $p$ expansion is discussed. Those textbooks remain at the level of introducing the coefficients $a_{k}$'s as in Eq. (\ref{bexpareal}) but do not discuss the associated digit function Eq. (\ref{cucuAreal}), whose properties we are about to exploit here. 

We first note that the digit function allows Cantor's proof of the fact that \emph{the real numbers cannot be counted with the natural numbers} to be presented in a concise way. If we concentrate on the real numbers in the unit interval $[0,1)$ one can show that, already these, cannot be counted with the natural numbers. For let us assume an infinite list of such numbers $r_{1}, r_{2},\ldots, r_{n}, \ldots$ where $n \in \mathbb{N}$. Then, the digit expansion in radix $p$, Eq. (\ref{idenreal}) of each such number is 
\begin{equation}
r_{n}=\sum_{k=-\infty}^{-1} p^{k} \mathbf{d}_{p}(k,r_{n})=\sum_{k=1}^{\infty} p^{-k} \mathbf{d}_{p}(-k,r_{n}) \label{cantordia1}
\end{equation}
Clearly, the number $x \in [0,1)$ given by
\begin{equation}
x=\sum_{k=1}^{\infty} p^{-k}  \mathbf{d}_{p}(0,\mathbf{d}_{p}(-k,r_{k})+1)  \label{cantordia2}
\end{equation}
is not in the above list since at least one digit of this number is different to the ones of each number $r_{n}$, $n=1,2 \ldots$ in the list. The existence of this number $x$ proves that no bijection between the naturals and the reals in the interval $[0,1)$ is possible and, hence, that the cardinal of $\mathbb{R}$ is larger than that of $\mathbb{N}$. This is the celebrated Cantor's diagonal argument and it finishes the proof.

Let us now briefly discuss some elementary properties of the digit function, that we shall need below: we can shift $x$ to $x+np^{k+1}$ (with $n$ being an integer number) and the function remains invariant
\begin{eqnarray}
\mathbf{d}_{p}(k,x+np^{k+1})&=&\left \lfloor \frac{x+np^{k+1}}{p^{k}} \right \rfloor-p\left \lfloor \frac{x+np^{k+1}}{p^{k+1}} \right \rfloor \nonumber \\
&=&\left \lfloor \frac{x}{p^{k}}+np \right \rfloor-p\left \lfloor \frac{x}{p^{k+1}}+n \right \rfloor \nonumber \\
&=&\left \lfloor \frac{x}{p^{k}} \right \rfloor+np-p\left \lfloor \frac{x}{p^{k+1}} \right \rfloor-np \nonumber \\
&=&
\left \lfloor \frac{x}{p^{k}} \right \rfloor-p\left \lfloor \frac{x}{p^{k+1}} \right \rfloor = \mathbf{d}_{p}(k,x) \label{pro1}
\end{eqnarray}
i.e. the digit function is $p^{k+1}$-periodic. We have also a scaling property that concerns multiplication of $x$ by any integer power of $p$
\begin{eqnarray}
\mathbf{d}_{p}(k,p^{m}x)&=&\left \lfloor \frac{p^{m}x}{p^{k}} \right \rfloor-p\left \lfloor \frac{p^{m}x}{p^{k+1}} \right \rfloor \nonumber \\
&=& \left \lfloor \frac{x}{p^{k-m}} \right \rfloor-p\left \lfloor \frac{x}{p^{k-m+1}} \right \rfloor \nonumber \\
&=&\mathbf{d}_{p}(k-m,x) \nonumber
\end{eqnarray} 

Since $\mathbf{d}_{p}(k,x)$ is an integer $0 \le \mathbf{d}_{p}(k,x) \le p-1$ we also trivially have 
\begin{equation}
\mathbf{d}_{p}(0,\mathbf{d}_{p}(k,x))=\mathbf{d}_{p}(k,x)-p\left \lfloor \frac{\mathbf{d}_{p}(k,x)}{p} \right \rfloor=\mathbf{d}_{p}(k,x) \label{pro1b}
\end{equation}
because $\left \lfloor \mathbf{d}_{p}(k,x)/p \right \rfloor=0$. Let $m \in \mathbb{Z}$. By Euclidean division, we can always find integers $a$ and $b$ such that  $m=ap+b$ and $b=\mathbf{d}_{p}(0,m)$ and, thus, we also have
\begin{equation}
\mathbf{d}_{p}(0,n+\mathbf{d}_{p}(0,m))=\mathbf{d}_{p}(0,n+ap+\mathbf{d}_{p}(0,m))=\mathbf{d}_{p}(0,n+m)  \label{pro2b}
\end{equation}
where Eq. (\ref{pro1}) has also been used. Another property, that follows directly from the definition, is the following
\begin{eqnarray}
\mathbf{d}_{np}(k,x)&=&\mathbf{d}_{p}\left(k, \frac{x}{n^{k}} \right)+p\mathbf{d}_{n}\left(k, \frac{x}{p^{k+1}} \right) \nonumber \\
&=&\mathbf{d}_{n}\left(k, \frac{x}{p^{k}} \right)+n\mathbf{d}_{p}\left(k, \frac{x}{n^{k+1}} \right)    \label{reldi1}
\end{eqnarray}


Because of the above properties, it is clear that if $n \in \mathbb{Z}$ is any integer $\mathbf{d}_{p}(k,n)$ is an integer in the interval $[0,p-1]$. The digit function for $k=0$, $\mathbf{d}_{p}(0,n)$, maps the set of the integers $\mathbb{Z}$ to the set of residue classes modulo $p$ denoted as $\mathbb{Z}/p\mathbb{Z}$: It extracts the remainder of $n$ under division by $p$. Thus the digit function is an homomorphism between both sets. For example, if $p=2$, $\mathbf{d}_{2}(0,n)=0$ if $n$ is even and $\mathbf{d}_{p}(0,n)=1$ if $n$ is odd. 

It is a trivial exercise to show that $\mathbf{d}_{p}(0,n)$ provides indeed a group homomorphism of the integers $\mathbb{Z}$ under ordinary addition to the finite cyclic group  $\mathbb{Z}/p\mathbb{Z}$ of the integers in the interval $[0,p-1]$ under addition modulo $p$. Let $m$ and $n$ be such integers. The group axioms are all satisfied. We have \emph{1) Closure}: The digit function $\mathbf{d}_{p}(0,n)$ yields the remainder of $m+n$ under division by $p$ which is, of course, an integer $\in [0, p-1]$ as well; \emph{2) Associative property}: we have 
\begin{eqnarray}
&&\mathbf{d}_{p}\left(0, m+\mathbf{d}_{p}\left(0, n+k \right)\right)\nonumber \\
&&=\mathbf{d}_{p}\left(0, m+n+k \right)=\mathbf{d}_{p}\left(0, \mathbf{d}_{p}\left(0, m+n \right)+k \right) \nonumber
\end{eqnarray}
where Eq. (\ref{pro2b}) has been used twice; \emph{3) Neutral element}: The neutral element is 0, as in the ordinary sum; \emph{4) Inverse element}: The inverse element of $m$ is $p-m$, since $\mathbf{d}_{p}\left(0, m+p-m \right)=0$.

Thus, the group axioms are satisfied. Furthermore the group is \emph{abelian} since 
$\mathbf{d}_{p}\left(0, m+n \right)=\mathbf{d}_{p}\left(0, n+m \right)$, and it is \emph{cyclic} because a single element generates the whole group. To show this latter statement let us just consider $n=1$ and operate repeatedly with it. We have, from Eqs. (\ref{pro1}) and (\ref{pro1b})
\begin{eqnarray}
&&\mathbf{d}_{p}\left(0, 1+1 \right)=2, \ \ldots, \ \mathbf{d}_{p}\left(0, 1+p-2 \right)=p-1, \nonumber \\
&&\mathbf{d}_{p}\left(0, 1+p-1 \right)=0,\quad  \mathbf{d}_{p}\left(0, 1+p \right)=1  \nonumber
\end{eqnarray} 
Thus, the orbit of the element $1$ has order $p$ equal to the one of the group, $1$ can be made the only generator of the group, and the group is cyclic. 

The result of operating on $m$ and $n$ with the digit function $\mathbf{d}_{p}(0,m+n)$ is given in the following table
\begin{center}
\begin{tabular}{c|cccccc}
$\ \mathbf{d}_{p}(0,m+n)  \ $ & $ 0 \ $ & $\ 1 \ $ & $\ 2 \ $ & $\   \ldots  \ $ & $ \ p-2 \ $ & $ \ p-1 \ $ \\
\hline
$\ 0 $ & $ 0 \ $ & $\ 1 \ $ & $\ 2 \ $ & $\   \ldots  \ $ & $ \ p-2 \ $ & $ \ p-1 \ $ \\
$\ 1 $ & $ 1 \ $ & $\ 2 \ $ & $\ 3 \ $ & $\   \ldots  \ $ & $ \ p-1 \ $ & $ \ 0 \ $ \\
$\ \ldots $ & $ \ldots \ $ & $\ \ldots \ $ & $ \ \ldots \ $ & $\   \ldots  \ $ & $ \ \ldots \ $ & $ \ \ldots \ $ \\
$\ p-2 $ & $ p-2 \ $ & $\ p-1 \ $ & $\ 0 \ $ & $\   \ldots  \ $ & $ \ p-4 \ $ & $ \ p-3 \ $ \\
$\ p-1 $ & $ p-1 \ $ & $\ 0 \ $ & $\ 1 \ $ & $\   \ldots  \ $ & $ \ p-3 \ $ & $ \ p-2 \ $ \\
\end{tabular}
\end{center}
where $m$ is listed in the rows and $n$ in the columns. The table gives the output of the operation $\mathbf{d}_{p}(0,m+n)$ and coincides with the Cayley table of the finite cyclic group $\mathbb{Z}/p\mathbb{Z}$ of order $p$ \cite{Gallian}. Specific examples are given by
\begin{center}
\begin{tabular}{c|cc}
$\ \ \mathbf{d}_{2}(0,m+n)  \ \ $ & $ \ 0  \ $ & $\ \  1 \ \ $ \\
\hline
$\ 0 \ $ & $\ 0 \ $ & $\  1 \ $ \\
$\ 1 \ $ & $\ 1 \ $ & $ \ 0 \ $ \\
\end{tabular}
\qquad 
\begin{tabular}{c|ccc}
$\ \mathbf{d}_{3}(0,m+n)  \ $ & $ 0 \ $ & $\ 1 \ $ & $\   2 \  $ \\
\hline
$\ 0 $ & $\ 0 \ $ & $\ 1 \ $ & $\ 2 \ $ \\
$\ 1 $ & $\ 1 \ $ & $\ 2 \ $ & $\  0 \ $ \\
$\ 2 $ & $\ 2 \ $ & $\ 0 \ $ & $ \ 1 \ $ \\
\end{tabular}
\end{center}

The above tables have the latin square property characteristic of finite groups and other  magmas (as loops and quasigroups \cite{Ore}). Because of the group being cyclic each row of outcomes is a cyclic permutation of the previous row. The abelian character of the group is easily seen through the symmetry of the tables regarding their main diagonal. Since all integers $\in [0,p-1]$ appear in each row we have the following remarkable fact
\begin{eqnarray}
\sum_{n=0}^{p-1}\mathbf{d}_{p}\left(0, m+n \right)&=&\sum_{j=0}^{p-1}j=\frac{p(p-1)}{2} \label{simplead}
\end{eqnarray}
since by fixing $m\in [0,p-1]$ the sum in Eq. (\ref{simplead}) runs over all possible values $n\in [0,p-1]$ an so do the outcomes. Note the insensitivity to $m$ of the sum in Eq. (\ref{simplead}):  The latin square is also a magic square and all rows and columns add to the same quantity $\frac{p(p-1)}{2}$ as given by Eq. (\ref{simplead}). We exploit this property in the next section in establishing our main result.

\section{The $p\lambda n$ fractal decomposition} \label{thedecon}

The digit function is a natural building block for fractal curves as we shall now explain. In general, as the classical example of the Cantor set shows, if we systematically exclude from the real line certain intervals, by repeating this process at all scales we end up with a fractal set. We now describe how a closely related method can be mathematically implemented in an efficient manner with help of the digit function.

\begin{figure*} 
\includegraphics[width=0.75\textwidth]{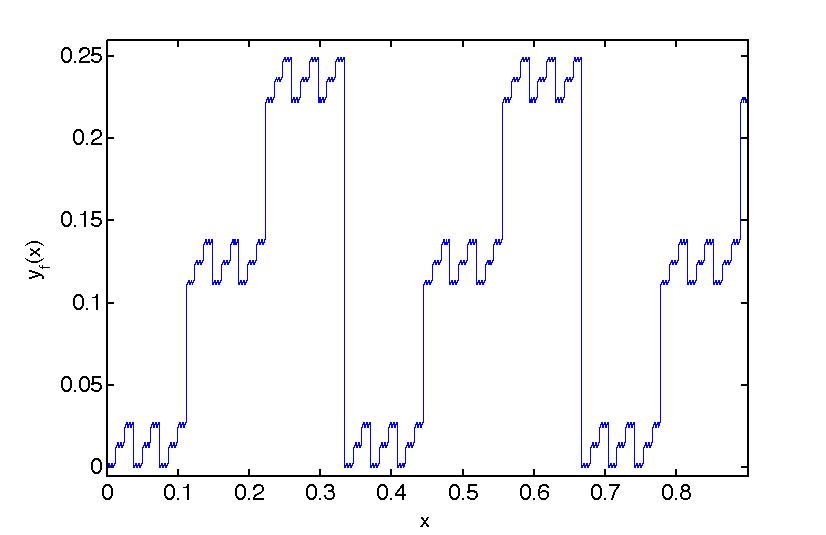}
\caption{The fractal curve $y_{f}(x)$ obtained from Eq. (\ref{idenf}) for $p=3$.} \label{fracty}
\end{figure*}

Let us, for example, set the digits for which $|k|$ is odd in Eq. (\ref{idenreal}) to zero. By means of the digit function this can be easily accomplished as
\begin{equation}
y_{f}(x)=\text{sign}(x)\sum_{k=-\infty}^{\lfloor \log_{p}|x| \rfloor} p^{k} \mathbf{d}_{p}(k,|x|)\mathbf{d}_{2}(0,|k|+1) \label{idenf}
\end{equation}
this function leads to the fractal curve shown in Fig. \ref{fracty} for $p=3$. Note that, when $k$ is odd, the terms $p^{k} \mathbf{d}_{p}(k,|x|)\mathbf{d}_{2}(0,|k|+1)$  do not contribute to the sum because $\mathbf{d}_{2}(0,k+1)=0$. In this way, we are systematically excluding intervals in the real line at all scales. The result is a self-similar discontinuous structure (had we all the digits we would obtain the line $y=x$). If we now consider the fractal curve where we set the digits with even $k$ to zero $\widehat{y}_{f}(x)=\text{sign}(x)\sum_{k=-\infty}^{\lfloor \log_{p}|x| \rfloor} p^{k} \mathbf{d}_{p}(k,|x|)\mathbf{d}_{2}(0,|k|)$ it is clear that the function $y(x)=x$ is equal to $y=y_{f}+\widehat{y}_{f}$ since $\mathbf{d}_{2}(0,|k|)+\mathbf{d}_{2}(0,|k|+1)=1$.

%

We have considered digits in even or odd positions. But, having in mind the results in the previous section, we can now generalize this as follows. Let $f(x)$, $f:\mathbb{R} \to \mathbb{R}$ be any real-valued function of $x$ (it does not need to be neither continuous nor measurable). We define the $p\lambda n$ fractal decomposition of $f(x)$  and denote it $\mathbf{F}_{p\lambda n}f(x)$ as
\begin{eqnarray}
&&\mathbf{F}_{p\lambda n}f(x)\equiv \label{lampm} \\
&&\equiv \frac{2\ \text{sign}f(x)}{\lambda(\lambda-1)}\sum_{k=-\infty}^{\lfloor \log_{p}|f(x)| \rfloor}p^{k}\mathbf{d}_{p}(k,|f(x)|)\mathbf{d}_{\lambda}(0,|k|+n)  \nonumber
\end{eqnarray}
This definition means that out of the function $f(x)$ we construct the function $\mathbf{F}_{p\lambda n}f(x)$ that we shall call a \emph{fractal object}. There is a total of $\lambda$ such fractal objects, each one indexed by $n \in [0, \lambda-1]$. 

The meaning of this definition shall now be substantiated by proving the following fact, which is the main result of this article
\begin{equation}
f(x)=\sum_{n=0}^{\lambda-1}\mathbf{F}_{p\lambda n}f(x) \label{theresult}
\end{equation}
i.e. \emph{the sum of all $\lambda$ fractal objects yields the function from which they constitute a decomposition}. Note that \emph{the function $f(x)$ being decomposed can have any arbitrary number of variables $f(x,y,\ldots)$ and any dependence on them and the decomposition is equally valid. The result holds in general.} It is straightforward to prove it. Note that, because of Eq. (\ref{simplead}) we have
\begin{equation}
\sum_{n=0}^{\lambda-1}\mathbf{d}_{\lambda}(0,|k|+n)=\frac{\lambda(\lambda-1)}{2} 
\end{equation}
Therefore 
\begin{eqnarray}
&&\sum_{n=0}^{\lambda-1}\mathbf{F}_{p\lambda n}f(x)=\sum_{n=0}^{\lambda-1}\frac{2\ \text{\text{sign}}f(x)}{\lambda(\lambda-1)} \times \\
&& \qquad \qquad \quad \times \sum_{k=-\infty}^{\lfloor \log_{p}|f(x)| \rfloor}p^{k}\mathbf{d}_{p}(k,|f(x)|)\mathbf{d}_{\lambda}(0,|k|+n)
\nonumber \\
&&=\text{\text{sign}}f(x)\sum_{k=-\infty}^{\lfloor \log_{p}|f(x)| \rfloor}p^{k}\mathbf{d}_{p}(k,|f(x)|)\times \nonumber \\
&&\qquad \qquad \qquad \qquad \quad \times \left(\frac{2}{\lambda(\lambda-1)}\sum_{n=0}^{\lambda-1}\mathbf{d}_{\lambda}(0,|k|+n)\right) \nonumber \\
&&=\text{\text{sign}}f(x)\sum_{k=-\infty}^{\lfloor \log_{p}|f(x)| \rfloor}p^{k}\mathbf{d}_{p}(k,|f(x)|)=f(x) \nonumber
\end{eqnarray}
where the radix expansion of $f(x)$, Eq. (\ref{idenreal}) (i.e. replacing $x$ by $f(x)$ in that equation) has been used.

\begin{figure*}
\includegraphics[width=0.9\textwidth]{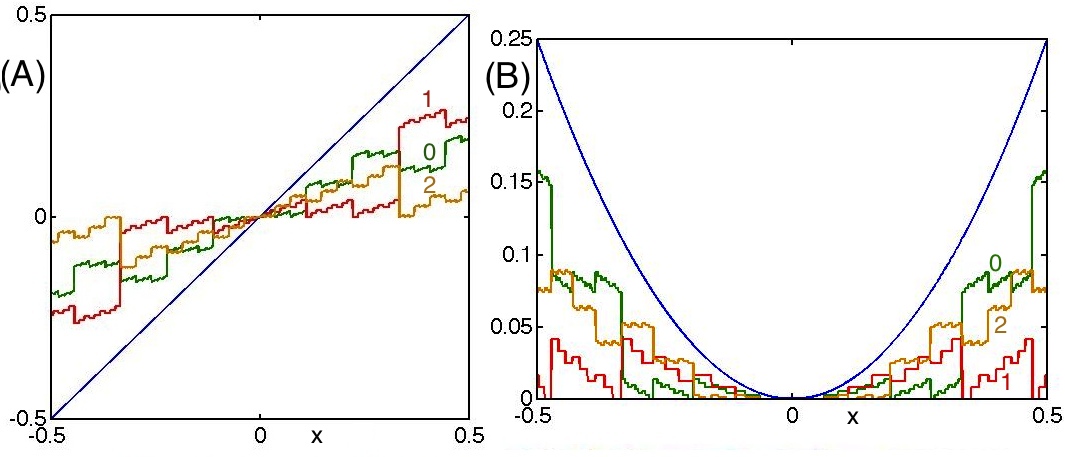}
\caption{(Color online) The continuous curves $f(x)=x$ (left) and $f(x)=x^{2}$ (right) and their associated fractal objects $\mathbf{F}_{3,3,n}f(x)$ (i.e. $p=3$, $\lambda=3$) for the values of $n$ indicated in the figure.} \label{fracty2}
\end{figure*}

In Fig. \ref{fracty2}  two simple examples of this result are shown. The functions $f(x)=x$ (left) and $f(x)=x^{2}$ (right) are plotted together with their fractal objects $\mathbf{F}_{3,3,n}f(x)$ (i.e. $p=3$, $\lambda=3$) calculated from Eq. (\ref{lampm}) for the values of $n$ indicated in the figure. One can see at a glance that the sum of the values of the fractal objects is equal to $f(x)$ for every value of $x$. The fractal objects cross themselves in nontrivial ways so that their sum is the continuous and differentiable function $f(x)$ in the figures. Note that, although the fractal objects are discontinuous, they are defined at every point. Since the limits from the right and from the left at the discontinuities are different, this requires that the jump of the fractal objects at such discontinuities agree so as to produce a continuous curve when added. This is warranted by the construction.


We have dealt with finite cyclic groups above. In general, let $g_{\lambda}(m,n)$ be a surjective application sending all integers $m, n$ to the integers $\in [0,\lambda-1]$. If the table of $g_{\lambda}(m,n)$ (of dimension $\lambda^{2}$) has the latin square property all integers $0,1,\ldots, p-1$ appear in every column and every row. Then, Eq. (\ref{simplead}) holds again and we have
\begin{eqnarray}
\sum_{n=0}^{\lambda-1}g_{\lambda}(m,n)&=&\sum_{j=0}^{\lambda-1}j=\frac{\lambda(\lambda-1)}{2} \label{simplead2}
\end{eqnarray}
and Eq. (\ref{lampm}) can thus be readily generalized as
\begin{eqnarray}
&&\mathbf{F}_{p\lambda n}f(x)\equiv \label{lampmg} \\
&&\equiv \frac{2\ \text{sign}f(x)}{\lambda(\lambda-1)}\sum_{k=-\infty}^{\lfloor \log_{p}|f(x)| \rfloor}p^{k}\mathbf{d}_{p}(k,|f(x)|)g_{\lambda}(|k|,n) \nonumber
\end{eqnarray}
and the main result Eq. (\ref{theresult}) trivially holds since the latin square property is all what is needed to prove that result. Thus, we can use \emph{all} finite groups at our disposal to generate fractal functions in a nontrivial but systematic way, provided that we have a construction for them through a knowledge of $g_{\lambda}(m,n)$. Indeed we have already found expressions for $g_{\lambda}(m,n)$ for all members of some infinite families of finite groups including all metacyclic groups (from which cyclic, dicyclic and dihedral groups are specific cases \cite{Coxeter2}), symmetric groups of permutations (which include all other finite groups as specific subgroups) and alternating groups. Work is in progress to find expressions for $g_{\lambda}(m,n)$ for Chevalley groups, which are specially interesting because they are the finite counterparts of Lie groups \cite{ConwayATLAS}. Since all these constructions have a purely mathematical character, they shall be presented elsewhere and we refrain from giving here more details.  We just give only one example of $g_{\lambda}(m,n)$ that describes the infinite family of dihedral groups $D_{2q}$ (i.e. the group of automorphisms of a regular polygon). These groups are noncommutative and have even order $\lambda=2q$ and we find
\begin{eqnarray}
g_{2q}(m,n)&=&
\mathbf{d}_{q}\left(0, m+(-1)^{\mathbf{d}_{2}\left(0,\frac{m}{q}\right)}n\right)+ \label{dihsum} \\
&&+q\mathbf{d}_{2}\left(0, \mathbf{d}_{2}\left(0,\frac{m}{q}\right)+\mathbf{d}_{2}\left(0,\frac{n}{q}\right)\right)  \nonumber
\end{eqnarray}  
For example, $q=3$ describes the group of rotations and reflections that fix an equilateral triangle, $q=4$ the ones that fix a square, $q=5$ a regular pentagon, etc. Setting $q$ to a given value, Eq. (\ref{dihsum}) automatically provides the corresponding Cayley table of the group. For example, for $q=3$ we obtain the Cayley table of $D_{6}$, given by the operation of $g_{6}(m,n)$ ($m$ is listed in the rows and $n$ in the columns)
\begin{center}
\begin{tabular}{c|cccccc}
$\  g_{6}(m,n) \ $ & $ 0 \ $ & $\ 1 \ $ & $\  2 \  $ & $\ 3 \ $ & $\ 4 \ $ & $\   5 \  $ \\
\hline
               $\ 0  \ $ & $\ 0 \ $ & $\ 1 \ $ & $\  2 \  $ & $\ 3 \ $ & $\ 4 \ $ & $\   5 \  $ \\
               $\ 1  \ $ & $\ 1 \ $ & $\ 2 \ $ & $\   0 \  $ & $\ 4 \ $ & $\ 5 \ $ & $\   3 \  $ \\
	       $\ 2  \ $ & $\ 2 \ $ & $\ 0 \ $ & $\   1 \  $ & $\ 5 \ $ & $\ 3 \ $ & $\  4 \  $ \\
	       $\ 3  \ $ & $\ 3 \ $ & $\ 5 \ $ & $\  4 \  $ & $\ 0 \ $ & $\ 2 \ $ & $\   1 \  $ \\
	       $\ 4  \ $ & $\ 4 \ $ & $\ 3 \ $ & $\   5 \  $ & $\ 1 \ $ & $\ 0 \ $ & $\  2 \  $ \\
	       $\ 5  \ $ & $\ 5 \ $ & $\ 4 \ $ & $\  3 \  $ & $\ 2 \ $ & $\ 1 \ $ & $\  0 \  $ \\
\end{tabular}
\end{center}
Thus, by using $g_{6}(m,n)$ in Eq. (\ref{lampmg}), we find the six different fractal objects $\mathbf{F}_{p6n}f(x)$ obtained from this dihedral group.

Although continuity and differentiability are generally lost through the action of the fractal decomposition, the fractal objects inherit any symmetry $\lambda(x)$ of the function $f(x)$. Thus if a transformation $x \to \lambda(x)$ is such that $f(\lambda(x))=f(x)$ then, we have $\mathbf{F}_{p\lambda n}f(\lambda(x))=\mathbf{F}_{p\lambda n}f(x)$ as well. For example, if $f(x)$ is periodic, i.e. $f(x+L)=f(x)$, so are also the fractal objects necessarily.

We also note that any permutation of the fractal objects $\mathbf{F}_{p\lambda n}f(x)$ through the label $n$, i.e. any bijection that sends $\mathbf{F}_{p\lambda n}f(x)$ to $\mathbf{F}_{p\lambda n'}f(x)$, where both $n$ and $n'$ are $\in [0,\lambda-1]$ leaves $f(x)$ invariant. Thus, $f(x)$ is invariant under the action of the symmetric group of permutations of order $\lambda$, $S_{\lambda}$, that exchanges the fractal objects in its decomposition.

We have introduced the fractal decomposition in terms of the operator $\mathbf{F}_{p\lambda n}$ acting on $f$. With this in mind, the result of the theorem can also be formally expressed as
\begin{equation}
\mathbf{1}=\sum_{m=0}^{\lambda-1}\mathbf{F}_{p\lambda n}
\end{equation}
where $\mathbf{1}$ is the identity operator. In this way no reference to a particular $f$ is made. One can easily show that the set of $\lambda$ operators $\mathbf{F}_{p\lambda n}$ forms a finite ring under the operations
\begin{eqnarray}
\mathbf{F}_{p\lambda n}*\mathbf{F}_{p\lambda n'}&\equiv&\mathbf{F}_{p\lambda (n+n')} \\
\mathbf{F}_{p\lambda n}\odot\mathbf{F}_{p\lambda n'}&\equiv&\mathbf{F}_{p\lambda (n\cdot n')}
\end{eqnarray}
If $\lambda$ is prime this ring becomes a finite Galois field $\mathbb{F}_{\lambda}$ of $\lambda$-characteristic under these operations. 

As in quantum mechanics the operators $\mathbf{F}_{p\lambda n}$ can also be made to act on vectors. Care must then be taken with their proper ordering, since these operators do not generally commute. For example, one can define the exponential operator $e=e^{\textbf{1}} \equiv e^{\sum_{m=0}^{\lambda-1}\mathbf{F}_{p\lambda n}}$ in terms of its Taylor series. Trivially, one has then, in general,
\begin{equation}
e^{\sum_{m=0}^{\lambda-1}\mathbf{F}_{p\lambda n}} \ne \prod_{m=0}^{\lambda-1}e^{\mathbf{F}_{p\lambda n}} \label{suzu}
\end{equation} 
i.e., one cannot factorize such operator because of the $\mathbf{F}_{p\lambda n}$ being non-commutative in general. This is reminiscent of quantum mechanics, where one finds a similar situation, leading to the Baker-Campbell-Hausdorff formula and the Trotter-Suzuki decomposition \cite{Suzuki1, Suzuki2, Suzuki3} which leads also to fractals \cite{Suzuki1}. 

We can then summarize our solution to the problem posed in the introduction: It is always possible to bring any Hamiltonian to the form of Eq. (\ref{ispos}). One then has
\begin{equation}
H_{n}=\mathbf{F}_{p\lambda n}H \label{finding}
\end{equation}
with $\mathbf{F}_{p\lambda n}H$ given by Eq. (\ref{lampm}). Or, even more generally, by Eq. (\ref{lampmg}).

\section{Extension to complex-valued functions}

\begin{figure*}
\includegraphics[width=1.0\textwidth]{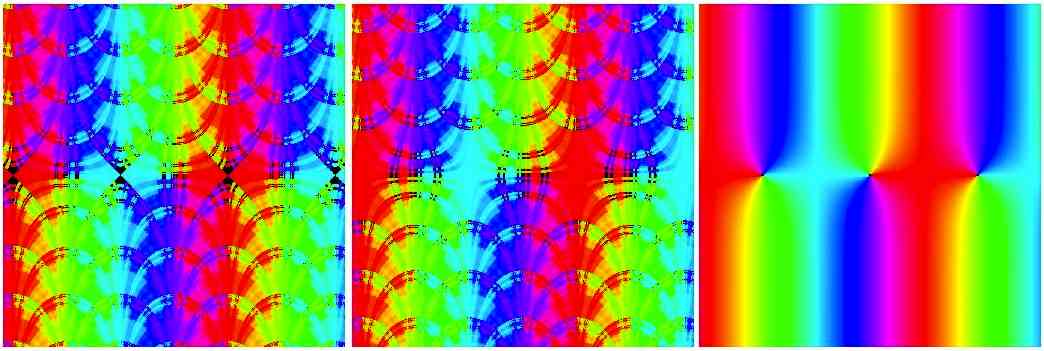}
\caption{\scriptsize{(Color online) Plot of the analytic complex function $f(z)=\sin z$  (right) and their associated fractal objects $\mathbf{F}_{2,2,n}f(z)$ (i.e. $p=2$, $\lambda=2$) for the values of $n=0$ (left) and $n=1$ (center). Isochromatic lines denote regions of constant phase. The modulus of $f(z)$ thus varies along isochromatic lines. The window displayed is $x \in [-5,5]$ (horizontal axis) and $y \in [-5,5]$ (vertical axis).}} \label{sinzfra}
\end{figure*}

Let $z=|z|e^{i\varphi}$ be a complex number with $\varphi$ denoting its phase and $|z|$ its modulus. We have that $z=x+iy$, with $x= (z+\overline{z})/2$ and $y= (z-\overline{z})/(2i)$ being real numbers ($\overline{z}$ denotes the complex conjugate of $z$) and it is clear that we can extend the definition of the digit function to the complex plane as follows
\begin{equation}
\mathbf{d}_{p}(k,z)\equiv \mathbf{d}_{p}(k,x)+i\mathbf{d}_{p}(k,y)
\end{equation}
The numbers $\mathbf{d}_{p}(k,x)$ and $\mathbf{d}_{p}(k,y)$ are non-negative integers in the interval $[0,p-1]$ the digit function $\mathbf{d}_{p}(k,z)$ is a Gaussian integer in the square with vertices $(0,\ 0)$, $(p-1,\ 0)$, $(0,\ p-1)$, $(p-1,\ p-1)$ (vertices included).

Since $|z| \ge \max\{ |x|, |y| \}$, we have
\begin{equation}
z=\sum_{k=-\infty}^{\lfloor \log_{p}|z| \rfloor} p^{k} \left(\text{sign}(x)\mathbf{d}_{p}(k,|x|)+i\ \text{sign}(y)\mathbf{d}_{p}(k,|y|)\right) \label{idencompl2}
\end{equation}
We define the \emph{complex digit function} $\mathbf{c}_{p}(k,z)$ as
\begin{equation}
\mathbf{c}_{p}(k,z)\equiv \text{sign}(x)\mathbf{d}_{p}(k,|x|)+i\ \text{sign}(y)\mathbf{d}_{p}(k,|y|)
\label{cdigit}
\end{equation}
so that Eq. (\ref{idencompl2}) is most concisely expressed as
\begin{equation}
z=\sum_{k=-\infty}^{\lfloor \log_{p}|z| \rfloor} p^{k} \mathbf{c}_{p}(k,z) \label{idencom}
\end{equation}
We note that $\mathbf{c}_{p}(k,z)$ yields a Gaussian integer in the square with vertices $(-p+1,\ -p+1)$, $(p-1,\ -p+1)$, $(-p+1,\ p-1)$, and $(p-1,\ p-1)$ (vertices included).
Since we also have
\begin{equation}
z=e^{i\varphi}|z|=e^{i\varphi}\sum_{k=-\infty}^{\lfloor \log_{p}|z| \rfloor} p^{k}\mathbf{d}_{p}(k,|z|) \label{idencompl3}
\end{equation}
the following identity holds
\begin{equation}
\mathbf{c}_{p}(k,z)=e^{i\varphi}\mathbf{d}_{p}(k,|z|)=\frac{z}{|z|}\mathbf{d}_{p}(k,|z|) \label{gaugeinv}
\end{equation}


With all these definitions and straightforward identities in mind, the fractal decomposition can now be easily extended to complex numbers in an analogous manner to the one described above for the reals. The $p\lambda n$ fractal decomposition of a complex function $f(z)$, $f:\mathbb{C}\to \mathbb{C}$, takes the form
\begin{eqnarray}
&&\mathbf{F}_{p\lambda n}f(z) \equiv \label{lamcompl} \\
&&\frac{2}{\lambda(\lambda-1)}\sum_{k=-\infty}^{\lfloor \log_{p}|f(z)| \rfloor}p^{k}\mathbf{c}_{p}(k,f(z))\mathbf{d}_{\lambda}(0,|k|+n)  \nonumber
\end{eqnarray}
and we have, as above
\begin{equation}
f(z)=\sum_{n=0}^{p-1}\mathbf{F}_{p\lambda n}f(z) \label{resulcom}
\end{equation}
and the fractal objects $\mathbf{F}_{p\lambda n}$ form again a Galois field $\mathbb{F}_{\lambda}$ if $\lambda$ is prime. As above, the factor $\mathbf{d}_{\lambda}(0,|k|+n)$ can be replaced in Eq. (\ref{lamcompl}) by an operator $g_{\lambda}(k,n)$ with the latin square property and Eq. (\ref{resulcom}) still holds (although the Galois field property is then generally lost and we have then a finite ring).

In Fig. (\ref{sinzfra}) the complex-valued fractal objects $\mathbf{F}_{2,2,n}\sin z$ (i.e. $p=2$, $\lambda=2$ and $f(z)=\sin z$) calculated from Eq. (\ref{lamcompl}) for the values of $n=0$ (left) and $n=1$ (center) are shown. The sum of these objects yields $f(z)=\sin z$ (right), as calculated from Eq. (\ref{resulcom}). We emphasize the fact that although the fractal objects generally fail to be analytic, the symmetries of $f(z)$ are also symmetries of the fractal objects. 
Thus, if for example $f(z)$ is invariant under a M\"obius transformation so that 
\begin{equation}
f\left(\frac{az+b}{cz+d}\right)=f(z)
\end{equation}
for some integers $a$, $b$, $c$, $d$, such that $ad-bc > 0$, then each fractal object in the decomposition is also invariant under the same M\"obius transformation.
In the case of $f(z)=\sin z$ in the figure, since the function exhibits a periodic behavior, the fractal objects have this overall periodicity as well and are somehow `phase locked' to $f(z)$ (albeit with a more detailed and fine structure). This is clearly noticed in the figure: Stripes with a certain color (the isochromatic regions denote constant phase regions) occupy the same positions in the fractal objects as in $f(z)$. Also, the zeros of $f(z)$ are zeros of the fractal objects as well (although the contrary is not true).

\section{The generalized $p\lambda n$ fractal decomposition}

A wide family of fractal decompositions, that include the $\mathbf{F}_{p\lambda n}$ decomposition as a specific case, can now be constructed by further inspecting the structure of Eq. (\ref{simplead}) which is instrumental to establish the decomposition, Eq. (\ref{theresult}). We state this result rigorously as a theorem, prove it, and give further concrete examples of fractal decompositions.

\noindent \textbf{Theorem.} \emph{Let $S$ be the set of integers $\in [0, \lambda-1]$ ($\lambda \ge 2,\ \lambda \in \mathbb{N}$) and let the digit function $\mathbf{d}_{\lambda}(k,x)$ for $x\in \mathbb {R}$ and $k\in \mathbb{Z}$ be defined by Eq. (\ref{cucuAreal}). Let also $g_{\lambda}(m,n): \mathbb{Z}\times \mathbb{Z} \to S$ be a surjective application such that $g_{\lambda}(m,n)=g_{\lambda}(\mathbf{d}_{\lambda}(0,m),\mathbf{d}_{\lambda}(0,n))$ is also a law of composition (magma) $S\times S \to S$ with the latin square property. Then, for any injective function $h: S \to \mathbb{R}^{+}$ (where $\mathbb{R}^{+}$ denote the non-negative real numbers) the following relationship holds}
\begin{equation}
\sum_{n=0}^{\lambda-1}h\left(g_{\lambda}(m,n)\right)=\sum_{j=0}^{\lambda-1}h\left(j\right)=\sigma_{\lambda} \label{magima}
\end{equation}
\emph{where $\sigma_{\lambda} \in \mathbb{R}^{+}$ is a nonzero constant, independent of $m$. Furthermore, for any function $f(x): \mathbb{R} \to \mathbb{R}$, there exist $\lambda$ distinct fractal objects $G_{p\lambda n}$ defined as}
\begin{eqnarray}
&&\mathbf{G}_{p\lambda n}f(x)\equiv \label{lolo} \\
&&\equiv  \frac{\text{sign}f(x)}{\sigma_{\lambda}}\sum_{k=-\infty}^{\lfloor \log_{p}|f(x)| \rfloor}p^{k}\mathbf{d}_{p}(k,|f(x)|) h\left(g_{\lambda}(|k|,n)\right) \nonumber 
\end{eqnarray}
\emph{that constitute a generalized $p\lambda n$ fractal decomposition, satisfying the following relationship}
\begin{equation}
f(x)=\sum_{n=0}^{\lambda-1}\mathbf{G}_{p\lambda n}f(x) \label{theresultgen}
\end{equation}
~\\
\noindent \emph{Proof:} The surjective application $g_{\lambda}(m,n)$ of any two integers $m\in \mathbb{Z}$ and $n\in \mathbb{Z}$ homomorphically maps the cartesian product $\mathbb{Z}\times \mathbb{Z}$ to the cartesian product $S\times S$, where $S$ is the set of integers in the interval $[0, \lambda-1]$ ($\lambda \in \mathbb{N}$), and returns an integer also in $S$, since $g_{\lambda}(m,n)=g_{\lambda}(\mathbf{d}_{\lambda}(0,m),\mathbf{d}_{\lambda}(0,n)) \in S$. Now, since the law of composition $g_{\lambda}(\mathbf{d}_{\lambda}(0,m),\mathbf{d}_{\lambda}(0,n))$ has the latin square property, all $\lambda$ distinct integers $j\in S$, $0\le j \le \lambda -1$ appear in each row $\mathbf{d}_{\lambda}(0,m)$ of its Cayley table. The injective function $h: S \to \mathbb{R}^{+}$ maps each of these distinct $\lambda$ integers to a \emph{distinct} non-negative real number $h\left(g_{\lambda}(\mathbf{d}_{\lambda}(0,m),\mathbf{d}_{\lambda}(0,n))\right)$ (distinctness is preserved because the function $h$ is injective). Therefore, the latin square $g_{\lambda}(\mathbf{d}_{\lambda}(0,m),\mathbf{d}_{\lambda}(0,n))$ with entries in $S$ becomes a latin square $h\left(g_{\lambda}(\mathbf{d}_{\lambda}(0,m),\mathbf{d}_{\lambda}(0,n))\right)$ with entries in $\mathbb{R}^{+}$. If we now fix $m$ and change $n$ so that $\mathbf{d}_{\lambda}(0,n)$ changes from $0$ to $\lambda-1$ it is clear that $h\left(g_{\lambda}(\mathbf{d}_{\lambda}(0,m),\mathbf{d}_{\lambda}(0,n))\right)$ ranges over all $\lambda$ distinct  non-negative images $h(j)$ of all distinct $\lambda$ entries $j\in S$ which are a permutation of all integers $\in S$. Each different $m$ leads to a different permutation $h(j')$ of the same distinct $\lambda$ values. Since all these permutations contain the same distinct elements in the image set, the sum of the latter is the same, independently of $m$. This proves Eq. (\ref{magima}).  From the latter, since all $h(j)$ are distinct and non-negative, $\sigma_{\lambda} \ne 0$ and we have
\begin{equation}
\frac{1}{\sigma_{\lambda}}\sum_{n=0}^{\lambda-1}h\left(g_{\lambda}(|k|,n)\right)=1\label{magimagi}
\end{equation}
for any $k \in \mathbb{Z}$. We now observe that, since each $h\left(g_{\lambda}(|k|,n)\right)$ is non-negative, $\left|\frac{h\left(g_{\lambda}(|k|,n)\right)}{\sigma_{\lambda}}\right| < 1$ and, hence, for any well defined function $f(x)$ expressed in radix $p$ as $f(x)=\text{\text{sign}}f(x)\sum_{k=-\infty}^{\lfloor \log_{p}|f(x)| \rfloor}p^{k}\mathbf{d}_{p}(k,|f(x)|)$ the fractal object $\mathbf{G}_{p\lambda n}f(x)$ given by Eq. (\ref{lolo}) is necessarily well defined as well, being Cauchy-convergent on each point $x$ to a certain real value (even when continuity and differentiability, which both depend on the specific forms of the functions $h$ and $f(x)$, is not generally warranted). Furthermore,
\begin{eqnarray}
&&\sum_{n=0}^{\lambda-1}\mathbf{G}_{p\lambda n}f(x)=\sum_{n=0}^{\lambda-1}\frac{\text{\text{sign}}f(x)}{\sigma_{\lambda}} \times \\
&& \qquad \qquad \quad \times \sum_{k=-\infty}^{\lfloor \log_{p}|f(x)| \rfloor}p^{k}\mathbf{d}_{p}(k,|f(x)|)h\left(g_{\lambda}(|k|,n)\right)
\nonumber \\
&&=\text{\text{sign}}f(x)\sum_{k=-\infty}^{\lfloor \log_{p}|f(x)| \rfloor}p^{k}\mathbf{d}_{p}(k,|f(x)|)\times \nonumber \\
&&\qquad \qquad \qquad \qquad \quad \times \left(\frac{1}{\sigma_{\lambda}}\sum_{n=0}^{\lambda-1}h\left(g_{\lambda}(|k|,n)\right)\right) \nonumber \\
&&=\text{\text{sign}}f(x)\sum_{k=-\infty}^{\lfloor \log_{p}|f(x)| \rfloor}p^{k}\mathbf{d}_{p}(k,|f(x)|)=f(x) \nonumber
\end{eqnarray}
where Eq. (\ref{magimagi}) has been used. This completes the proof of the theorem. $\Box$
~\\

The theorem also holds for a function of several real or complex variables $f(z, w, \ldots)$ because no property of $f(x)$ is used in proving the theorem other than the fact that it admits a radix-$p$ expansion. Functions with an arbitrary number of variables also generally admit such expansion. Also, any complex-valued function $f(z, w,\ldots)$ of an arbitrary number of variables, admits a radix-$p$ expansion by means of the complex digit function $\mathbf{c}_{p}(k,f(z, w,\ldots))$. In this latter case one has
\begin{eqnarray}
&&\mathbf{G}_{p\lambda n}f(z, w,\ldots)\equiv \label{loloz} \\
&&\equiv  \frac{1}{\sigma_{\lambda}}\sum_{k=-\infty}^{\lfloor \log_{p}|f(z, w,\ldots)| \rfloor}p^{k}\mathbf{c}_{p}(k,f(z, w,\ldots)) h\left(g_{\lambda}(|k|,n)\right) \nonumber 
\end{eqnarray}
We then have,
\begin{equation}
f(z, w,\ldots)=\sum_{n=0}^{\lambda-1}\mathbf{G}_{p\lambda n}f(z, w,\ldots) \nonumber
\end{equation}

\noindent \textbf{Corollary.} \emph{Let $\lambda_{0}, \lambda_{1}, \ldots, \lambda_{M}$ be natural numbers, all $\ge 2$ and let $f$ be an arbitrary function that admits a radix-$p$ expansion. Then,}
\begin{equation}
f=\sum_{n_{0}=0}^{\lambda_{1}-1}\sum_{n_{2}=0}^{\lambda_{2}-1}\ldots \sum_{n_{M}=0}^{\lambda_{M}-1}\mathbf{G}_{p\lambda_{M} n_{M}}\ldots \mathbf{G}_{p\lambda_{2} n_{2}}\mathbf{G}_{p\lambda_{1} n_{1}}f \label{composer}
\end{equation}

\noindent \emph{Proof:} Since $f$ admits a radix-$p$ expansion, so does it $f_{1}\equiv\mathbf{G}_{p\lambda_{1} n_{1}}f$ as well, as a result of the theorem, and so does $f_{2}\equiv\mathbf{G}_{p\lambda_{2} n_{2}}f_{1}=\mathbf{G}_{p\lambda_{2} n_{2}}\mathbf{G}_{p\lambda_{1} n_{1}}f$. One then has that $f_{M}\equiv\mathbf{G}_{p\lambda_{M} n_{M}}f_{M-1}=\mathbf{G}_{p\lambda_{M} n_{M}}\mathbf{G}_{p\lambda_{M-1} n_{M-1}}\ldots \mathbf{G}_{p\lambda_{1} n_{1}}f$ is also well defined. We now note
\begin{equation}
\sum_{n_{M}=0}^{\lambda_{M}-1}f_{M}=\sum_{n_{M}=0}^{\lambda_{M}-1}\mathbf{G}_{p\lambda_{M} n_{M}}f_{M-1}=f_{M-1}
\end{equation}
from Eq. (\ref{theresultgen}). Therefore,
\begin{equation}
\sum_{n_{M-1}=0}^{\lambda_{M-1}-1}\sum_{n_{M}=0}^{\lambda_{M}-1}f_{M}=\sum_{n_{M-1}=0}^{\lambda_{M-1}-1}f_{M-1}=f_{M-2}
\end{equation}
also from Eq. (\ref{theresultgen}) and we have, as well (by induction) 
\begin{equation}
\sum_{n_{1}=0}^{\lambda_{1}-1}\ldots\sum_{n_{M-1}=0}^{\lambda_{M-1}-1}\sum_{n_{M}=0}^{\lambda_{M}-1}f_{M}=\sum_{n_{1}=0}^{\lambda_{1}-1}f_{1}=f 
\end{equation}
which is Eq. (\ref{composer}). $\Box$

We now give some examples of the theorem. In all cases $h(j)$ constitutes an example of a suitable injective function. ~\\

\noindent \emph{Example 1:} If $h(j)=j$ ($j \in S$), and $g_{\lambda}(m,n)=\mathbf{d}_{\lambda}(0,m+n)$ we have $\sigma_{\lambda}=\frac{\lambda(\lambda-1)}{2}$ in Eq. (\ref{magima}) and
 $\mathbf{G}_{p\lambda n}$ given by Eq. (\ref{lolo}) coincides with $\mathbf{F}_{p\lambda n}$ given by Eq. (\ref{lampm}).
~\\

\noindent \emph{Example 2:} If $h(j)=\ln(1+j)$, with $j \in S$ and $g_{\lambda}(m,n)=\mathbf{d}_{\lambda}(0,m+n)$ we have $\sigma_{\lambda}=\ln \lambda!$ in Eq. (\ref{magima}) and $\mathbf{G}_{p\lambda n}$ given by Eq. (\ref{lolo}) takes the form
\begin{eqnarray}
&&\mathbf{G}_{p\lambda n}f(x)\equiv  \frac{\text{sign}f(x)}{\ln \lambda!} \times \label{lola2} \\
&&\qquad \times \sum_{k=-\infty}^{\lfloor \log_{p}|f(x)| \rfloor}p^{k}\mathbf{d}_{p}(k,|f(x)|) \ln \left(1+\mathbf{d}_{\lambda}(0,|k|+n)\right) \nonumber
\end{eqnarray}

\noindent \emph{Example 3:} If $h(j)=j^{q}$, with $j \in S$ and $q\in \mathbb{N}$ and $g_{\lambda}(m,n)=\mathbf{d}_{\lambda}(0,m+n)$ we have, by the Seki-Bernoulli formula \cite{Arakawa}, that 
\begin{equation}
\sigma_{\lambda}=\sum_{\ell=0}^{q}{q \choose \ell}\frac{(\lambda-1)^{q+1-\ell}}{q+1-\ell} B_{\ell} \label{Seki}
\end{equation}
in Eq. (\ref{magima}). Here $B_{\ell}$ are the Bernoulli numbers $B_{0}=1$, $B_{1}=\frac{1}{2}$, $B_{2}=\frac{1}{6}$, $B_{3}=0$, $B_{4}=-\frac{1}{30}$, etc. Then $\mathbf{G}_{p\lambda n}$ given by Eq. (\ref{lolo}), takes the form
\begin{eqnarray}
&&\mathbf{G}_{p\lambda n}f(x)\equiv  \frac{\text{sign}f(x)}{\sum_{\ell=0}^{q}{q \choose \ell}\frac{(\lambda-1)^{q+1-\ell}}{q+1-\ell}B_{\ell}} \times \label{lola3} \\
&&\qquad \times \sum_{k=-\infty}^{\lfloor \log_{p}|f(x)| \rfloor}p^{k}\mathbf{d}_{p}(k,|f(x)|) \left[\mathbf{d}_{\lambda}(0,|k|+n)\right]^{q} \nonumber
\end{eqnarray}
Let us apply this to the case $\lambda=4$ and $q=3$ to better illustrate this example and the proof of the theorem. We start with the latin square $g_{4}(m,n)=\mathbf{d}_{4}(0,m+n)$
\begin{center}
\begin{tabular}{c|cccc}
$\ \mathbf{d}_{4}(0,m+n)  \ $ & $ 0 \ $ & $\ 1 \ $ & $\   2 \ $ & $ \ 3 \ $\\
\hline
$\ 0 $ & $\ 0 \ $ & $\ 1 \ $ & $\ 2 \ $ & $\ 3 \ $\\
$\ 1 $ & $\ 1 \ $ & $\ 2 \ $ & $\  3 \ $ & $\ 0 \ $\\
$\ 2 $ & $\ 2 \ $ & $\ 3 \ $ & $ \ 0 \ $ & $\ 1 \ $\\
$\ 3 $ & $\ 3 \ $ & $\ 0 \ $ & $ \ 1 \ $ & $\ 2 \ $\\
\end{tabular}
\end{center}
with $m$ listed in the rows and $n$ in the columns, for values in the set $S$. Now, this latin square is sent by the injective application $h=j^{3}$ to another latin square
\begin{center}
\begin{tabular}{c|cccc}
$\ \left[\mathbf{d}_{4}(0,m+n)\right]^{3}  \ $ & $ 0 \ $ & $\ 1 \ $ & $\   2 \ $ & $ \ 3 \ $\\
\hline
$\ 0 $ & $\ 0 \ $ & $\ 1 \ $ & $\ 8 \ $ & $\ 27 \ $\\
$\ 1 $ & $\ 1 \ $ & $\ 8 \ $ & $\  27 \ $ & $\ 0 \ $\\
$\ 2 $ & $\ 8 \ $ & $\ 27 \ $ & $ \ 0 \ $ & $\ 1 \ $\\
$\ 3 $ & $\ 27 \ $ & $\ 0 \ $ & $ \ 1 \ $ & $\ 8 \ $\\
\end{tabular}
\end{center}
The sum of any row or any column of this latin square is $\sigma_{4}=36$ as predicted by the Seki-Bernoulli formula, Eq. (\ref{Seki}), for $\lambda=4$ and $q=3$,
\begin{eqnarray}
\sigma_{4}&=&\sum_{\ell=0}^{3}{3 \choose \ell}\frac{3^{4-\ell}}{4-\ell} B_{\ell} \nonumber \\
&=& {3 \choose 0}\frac{3^{4}}{4}+{3 \choose 1}\frac{3^{3}}{3}\frac{1}{2}+{3 \choose 2}\frac{3^{2}}{2}\frac{1}{6} + {3 \choose 3}\frac{3^{1}}{1}\cdot 0 \nonumber \\
&=& 36 \nonumber
\end{eqnarray}
Therefore, Eq. (\ref{lola3}) reduces in this specific case to
\begin{eqnarray}
&&\mathbf{G}_{p4 n}f(x)\equiv  \frac{\text{sign}f(x)}{36} \times \label{lola4} \\
&&\qquad \times \sum_{k=-\infty}^{\lfloor \log_{p}|f(x)| \rfloor}p^{k}\mathbf{d}_{p}(k,|f(x)|) \left[\mathbf{d}_{4}(0,|k|+n)\right]^{3} \nonumber
\end{eqnarray}

\noindent \emph{Example 4:} If $h(j)=r^{j}$, with $j \in S$ and $r>0$, $r\in \mathbb{R}^{+}$  and $g_{\lambda}(m,n)=\mathbf{d}_{\lambda}(0,m+n)$ we have a geometric series and, hence, 
\begin{equation}
\sigma_{\lambda}=\frac{r^{\lambda}-1}{r-1}
\end{equation}
in Eq. (\ref{magima}). Therefore, from Eq. (\ref{lolo}), 
\begin{eqnarray}
&&\mathbf{G}_{p\lambda n}f(x)\equiv  \frac{(r-1)\text{sign}f(x)}{r^{\lambda}-1} \times \label{lola5} \\
&&\qquad \times \sum_{k=-\infty}^{\lfloor \log_{p}|f(x)| \rfloor}p^{k}\mathbf{d}_{p}(k,|f(x)|) r^{\mathbf{d}_{\lambda}(0,|k|+n)} \nonumber
\end{eqnarray}

\section{Connection to statistical mechanics}

A formal connection can be established between digit expansions in positional number systems and the statistical mechanics of systems with a discrete energy spectrum. In this section we sketch this connection. We focus on non-negative real valued functions $f(x,y,\ldots) \ge 0$. The radix-$p$ expansion of such a function is given by Eq. (\ref{idenreal}) as
\begin{eqnarray}
f&=&\sum_{k=-\infty}^{\lfloor \log_{p}f \rfloor}p^{k}\mathbf{d}_{p}(k,f)  =\sum_{k=-\lfloor \log_{p}f \rfloor}^{\infty}p^{-k}\mathbf{d}_{p}(-k,f) \nonumber \\
 \label{idenfx}
\end{eqnarray}
Since $f \ge 0$ we can formally define a `Helmholtz free energy' $F$ such that $f=e^{-\beta F}$ with $\beta=1/(k_{B}T)$ denoting a Lagrange parameter, with $T$ the `temperature' and $k_{B}$ being `Boltzmann's constant'. We can then consider a `statistical system' with a discrete `energy' spectrum described by a `canonical ensemble' and where $E$ denotes energy, $E_{0}$ the energy of the ground state, $\epsilon$ a `characteristic energy' (governing the spacing of the energy levels), and $\Omega(E,n,v)$ the number of microstates with energy $E$, number of particles $n$, volume $v$ (other extensive variables that may also be present). If we introduce the following correspondence
\begin{eqnarray}
n &\equiv& x \qquad v \equiv y \qquad \text{etc.} \label{nparti} \\
e^{-\beta F} &\equiv& f(x,y) \label{freeE} \\
\frac{E}{\epsilon} &\equiv& k  \label{co1}\\
\frac{E_{0}}{\epsilon} &\equiv& -\lfloor \log_{p}f(x,y) \rfloor \label{co2} \\
\frac{\epsilon}{k_{B}T} &\equiv& \beta \epsilon \equiv \ln p \label{co3} \\
\Omega(E,n,v) &\equiv& \mathbf{d}_{p}(-k,f(x,y)) \label{co4}
\end{eqnarray}
then Eq. (\ref{idenfx}) becomes
\begin{equation}
e^{-\beta F}= \sum_{E=E_{0}}^{\infty}  \Omega(E,n,v) e^{-\beta E} \label{bwq2b}
\end{equation}
which indeed constitutes the definition of the canonical partition function $Z$ of the statistical system. It is then interesting to note that the change of radix $p$ in the number theoretical description of a non-negative real number mathematically mirrors the effect of a temperature change in the statistical system. Indeed, from the asymptotic limit $p\to \infty$ of Eq. (\ref{idenfx}), (cf. Eqs. (\ref{idenasi}) and (\ref{thermolim}))
\begin{eqnarray}
f&=&\lim_{p\to \infty}\sum_{k=-\lfloor \log_{p}f \rfloor}^{\infty}\mathbf{d}_{p}(-k,f) p^{-k}
\nonumber \\
&=&\mathbf{d}_{\infty}(0,f)
 \label{idenflim}
\end{eqnarray}
we find that $f$ equals its own most significative digit when written in an infinitely large radix (i.e. there is only one digit in this representation: the most representative one). This limit, because of Eq. (\ref{co3}), is similar to taking in the `statistical system' the limit $T \to 0$, so that the system is fully described by the ground energy state $E_{0}$. A simple example may help to clarify this. The positive number $f=e^{-\beta F}=204$ (e.g.) has only one digit in radix $p\ge 205$ (in radix $p=205$ one has $204=204\cdot 205^{0}$ and hence, only one power of 205 is needed to describe the number). In the decimal radix $p=10$, we have that $204=2\cdot 10^{2}+0\cdot 10^{1}+4\cdot 10^{0}$ has three digits, $2$, $0$ and $4$, and three powers of ten are needed. Finally, $204$ in radix $p=2$ has 8 digits, since $204=1\cdot 2^{7}+1\cdot 2^{6}+0\cdot 2^{5}+0\cdot 2^{4}+1\cdot 2^{3}+1\cdot 2^{2}+0\cdot 2^{1}+0\cdot 2^{0}$. Therefore, if we lower the radix we need more digits to describe the same number. The statistical correspondence established above shows, through Eq. (\ref{co3}), that lowering the radix $p$ is equivalent to increasing the temperature $T$ in the statistical system at constant $Z=e^{-\beta F}$: In the low temperature limit, the system is found in only one energy level, i.e. the ground state; as temperature is increased, excited energy levels become occupied and more energy levels are needed to describe the state of the system. \emph{This increased number of energy levels of the statistical system mirrors the increased number of digits needed at lower radix to represent the number $e^{-\beta F}$}. 

We should now clarify how the $p\lambda n$ fractal decomposition enters in this discussion. To understand this let us consider the possibility that the number of particles in the system $n$ can fluctuate so that $n$ can range from $0$ to a maximum number $N_{max}$ of particles. The system is in contact with a reservoir at chemical potential $\mu$ with which it exchanges particles. A suitable, natural description of this system is provided in statistical thermodynamics by the grand-canonical potential $\Xi$ \cite{Pathria}. We have
\begin{equation}
e^{-\beta \Xi}=\sum_{n=0}^{N_{max}}e^{-\beta F(n)}e^{\beta \mu n} \label{grandca}
\end{equation}
where $F(n)$ is the free energy of a cluster of $n$ particles. This expression defines the grand-canonical partition function. From here we observe that if we identify $N_{max}$ with $\lambda-1$ where $\lambda$ is the number entering in a generalized $p\lambda n$ decomposition, we then note that we can further identify
\begin{equation}
e^{-\beta F(n)} e^{\beta \mu n} \equiv \mathbf{G}_{p\lambda n}e^{-\beta \Xi}
\end{equation}
and then Eq. (\ref{grandca}) becomes
\begin{equation}
e^{-\beta \Xi}=\sum_{n=0}^{\lambda-1}\mathbf{G}_{p\lambda n}e^{-\beta \Xi} \label{grandcano}
\end{equation}
which is consistent with Eq. (\ref{theresultgen}). We then note that the probability $\mathcal{P}(n)$ of observing a cluster with $n$ particles is given by
\begin{equation}
\mathcal{P}(n)=\frac{e^{-\beta F(n)}e^{\beta \mu n}}{e^{-\beta \Xi}}=e^{\beta \Xi}\mathbf{G}_{p\lambda n}e^{-\beta \Xi} \label{probabob}
\end{equation}
This expression is remarkable, since it is valid also if $\beta\Xi$ is complex-valued. It yields, in any case, a \emph{real-valued} probability. To see this note that if we make the transformation $\beta\Xi \to \beta\Xi+i\alpha$ (with $\alpha$ an arbitrary real number) we have, because of Eqs. (\ref{gaugeinv}) and Eq. (\ref{loloz}),
\begin{equation}
\mathbf{G}_{p\lambda n}\left(e^{-\beta \Xi-i\alpha}\right)=e^{-i\alpha}\mathbf{G}_{p\lambda n}e^{-\beta \Xi}
\end{equation}
Therefore, by introducing the same transformation $\Xi \to \Xi+i\alpha$ in Eq. (\ref{probabob}) we see that the latter remains invariant
\begin{equation}
\mathcal{P}(n)=e^{\beta \Xi+i\alpha}\mathbf{G}_{p\lambda n}e^{-\beta \Xi-i\alpha}=e^{\beta \Xi}\mathbf{G}_{p\lambda n}e^{-\beta \Xi}
\end{equation}
We have thus proved the phase invariance of Eq. (\ref{probabob}) and, thus, our methods are also able to describe Hilbert spaces, with Eq. (\ref{probabob}) giving the observable probability (Born rule) of a quantum state governed by a `wave function' $e^{-\beta \Xi}$ where $\beta \Xi$ can be an arbitrary complex number. Indeed, in quantum mechanics $\beta \Xi$ can be thought as having complex part equal to $iS/\hbar$ where $S$ is the Lagrangian action and $\hbar$ the Planck constant. We observe also the remarkable fact that the low-temperature limit $\beta \to \infty$ of a statistical system admits a dual interpretation under this connection as the limit $S/\hbar \to \infty$ of a quantum system. Thus, if we take the radix $p=\lfloor S/h \rfloor\equiv \eta \to \infty$ \cite{QUANTUM} the 'ground state' of the statistical system is nothing but the classical limit of a physical system where no nontrivial partitions are observed (note that in this regime, Eq. (\ref{thermolim}) holds).  In fact, we explained in \cite{QUANTUM} how, when demanding that the radix $\eta$ has the least economy, physical laws can be derived in a unified manner. 

Coming back to our connection to statistical mechanics, If we now further consider a situation where the volume of the system $V=\mathcal{N}v$ can also fluctuate in multiples $\mathcal{N}$ of an specific volume $v$ at the pressure of the reservoir $p$ we now have that this situation is well defined by a generalized free energy (subdivision potential) $\mathcal{E}$ \cite{HillBOOK, Chamberlin, Chamberlin3} as
\begin{equation}
e^{-\beta \mathcal{E}}=\sum_{\mathcal{N}=0}^{\mathcal{N}_{max}}e^{-\beta \Xi(\mathcal{N})}e^{-\beta pv \mathcal{N}} \label{grandcajen}
\end{equation}
This relationship defines the partition function for the generalized ensemble.
We find, again, by introducing $\lambda'=\mathcal{N}_{max}$ that we can identify
\begin{equation}
e^{-\beta \Xi(\mathcal{N})}e^{-\beta pv \mathcal{N}} \equiv \mathbf{G}_{p\lambda' \mathcal{N}}e^{-\beta \mathcal{E}}
\end{equation}
so that, again, from Eq. (\ref{grandcajen}) the consistent relationship $e^{-\beta \mathcal{E}}=\sum_{n=0}^{\lambda-1}\mathbf{G}_{p\lambda' \mathcal{N}}e^{-\beta \mathcal{E}}$ is warranted. The generalized ensemble describes small systems out of the thermodynamic limit
\cite{HillBOOK, Chamberlin, Chamberlin3} and has been successfully used by Chamberlin to derive a model for ferromagnetism \cite{Chamberlin} and a mean-field mesoscopic Landau theory \cite{Chamberlin3}. We note that the subdivision potential $\mathcal{E}$ is equal to zero in the thermodynamic limit as a consequence of the Gibbs-Duhem relationship, which precludes that all extensive variables in the system simultaneously fluctuate. Following previous work by Vives and Planes \cite{Vives} we found in \cite{HillTsallis} that the subdivision potential $\mathcal{E}$ of Hill's Nanothermodynamics can be connected with the Tsallis entropic form \cite{Tsallis}
\begin{equation}
S_{q}=\frac{\sum_{j}p_{j}^{q}-1}{1-q} \label{Tsallis}
\end{equation}
Here $q\in \mathbb{R}$ is the entropic parameter and $p_{j}$ is the probability of a microstate $j$. The interest of Tsallis entropy lies in its connection with (multi)fractals and complex systems \cite{Tsallis, VGM4, VGMStat} being a most prominent form of superstatistics \cite{Beck}. The connection found in \cite{HillTsallis} establishes that
\begin{equation}
\mathcal{E}=\frac{k_{B}T}{1-q}\left(1+\frac{1-q}{k_{B}}S_{q}\right)\ln \left(1+\frac{1-q}{k_{B}}S_{q}\right)-TS_{q}
\end{equation}
The entropic parameter $q$ controls the `distance' to the thermodynamic limit, which is approached when $q\to 1$. In that limit, by using L'Hopital's rule we observe from this expression that, indeed, $\mathcal{E}=0$, as mentioned above.

\section{Conclusions}

In this article we have established a general mathematical mechanism to partition any quantity into a finite set of fractal functions whose ordinary sum is equal everywhere to the original quantity. We have shown that there are infinite families of possible partitions indeed, all them exact, and we have explained how these partitions can be systematically constructed. These results have also been extended to complex-valued functions. 

The decomposition thus allows a finite family of fractal objects to be derived from an arbitrary `mother' function $f(z)$, which can be continuous and analytic. Perhaps a most interesting aspect of the construction is that, although the fractal objects generally fail to be continuous and analytic, they possess all internal symmetries of $f(z)$. 

A formal connection between our mathematical approach and statistical mechanics has also been briefly sketched. The interest of this connection, which shall be further explored elsewhere in the context of specific applications, lies in the fact that it provides a number-theoretic foundation for statistical mechanics. We have shown that digit expansions in positional number systems are naturally connected to partition functions as the ones arising in statistical mechanics. Interestingly, the radix change in the representation of numbers mirrors the effect of a temperature change in a statistical system. The connection also constitutes a pathway to the investigation of some certain contemporary approaches to statistical thermodynamics in which fractality, complexity and cooperativity play a central role \cite{HillBOOK, Chamberlin, Chamberlin3, Tsallis, TsallisPNAS, Beck, Beck2, TsallisSPECTRAL, VGMNano, VGMStat, PNAS2}. 

The radix representation of numbers is a topic that seems to have been neglected in Physics because of the obvious fact that a number is `always the same' independently of the radix used. Nonetheless, we have reasoned here why we do believe that the radix of numbers has an important physical meaning. \emph{Although a number is itself invariant under a radix change, its specific partitions in powers of the radix are not: A number, as represented in a standard positional number system, is also a partition function, where digits accompany powers of the radix, the total number of such powers being dependent both on the number and on the radix.} Therefore, the change of the radix of physical numbers has a physical impact because it governs the structure of the partitions of the number and the latter have a major importance in physics when one considers the quantum regime, stochastic phenomena, or the behavior of systems with many degrees of freedom. In a recent article \cite{QUANTUM}, we have identified a physical radix that governs the transition between classical and quantum mechanics and which is related to the Lagrangian action. When this radix is optimal, in a sense that it minimizes the size of a certain matrix called radix economy (or digit capacity), physical laws are derived in a straightforward manner and the principle of least action of classical mechanics is encompassed as well as the Hilbert space of quantum mechanics. In \cite{QUANTUM} we have also established a relationship between the optimal radix (given by the Lagrangian action) and Boltzmann entropy (see also \cite{Annals}) and in the present paper we have sketched how a change of the radix is related to a change of temperature in a canonical ensemble.

We believe that the approach presented here makes fractals amenable to theoretical investigation from a new perspective, in connection with problems involving quantization \cite{Kroeger, Suzuki1, Suzuki2, Suzuki3}. In quantum mechanics, symmetries are related to conservation laws that dictate which kind of physical processes at a fundamental level are possible, constraining them. Such symmetries and constraints are subtly and beautifully related through group theory. This article also establishes a general framework that connects fractals and finite group theory in a way that seems natural, and we hope that it may prove a useful tool for applications in both statistical mechanics and quantum field theory.


\bibliography{biblos}{}
\bibliographystyle{h-physrev3.bst}

\end{document}